\documentclass[english,keywords,amsmath,amssymb]{revtex4}
\usepackage[T1]{fontenc}
\usepackage[latin1]{inputenc}
\usepackage{babel}
\usepackage{graphicx}
\usepackage{color}
\usepackage{bm}
\usepackage{longtable}
\usepackage{amsmath} 
\usepackage{amsfonts}
\usepackage{amssymb}
\usepackage{hyperref}
\begin{document}
\title{Quantum contextuality for a three-level system sans realist model}
\author{A. K. Pan \footnote{akp@nitp.ac.in}}
\author{K. Mandal\footnote{kathakai.ism@gmail.com}}
\affiliation{ National Institute of Technology Patna, Ashok Rajhpath, Bihar 800005, India}

\begin{abstract}
  Recently, an interesting form of non-classical effect which can be considered as a form of contextuality within quantum mechanics, has been demonstrated for a  four-level system by discriminating the different routes that are taken for measuring a single observable. In this paper, we  provide a simpler version of that proof for a single qutrit, which is also within the formalism of quantum mechanics and without recourse to any realist hidden variable model. The degeneracy of the eigenvalues and the  L$\ddot{u}$der projection rule play important role in our proof. 
\end{abstract}
\maketitle
\section{introduction}
Quantum mechanics(QM) is a statistical theory and the occurrence of a definite outcome in an individual measurement of a dynamical variable cannot be ensured within the formalism. The realist hidden variable models of QM aim to provide the complete state of a quantum system by including hidden variables along with the quantum wave function. In such hypothetical models, the individual outcomes of a dynamical variable are desired to be predetermined by the suitable values of the hidden variables(say, $\lambda$s). In his path-breaking work, Bell \cite{bell64} first showed that in order to reproduce all QM statistics by a realist model, a constraint needs to be imposed at the level of individual measured value - a concept, widely known as nonlocality. Another similar constraint on the hidden variable model known as contextuality, was first put forward independently by  Bell \cite{bell} and by Kochen and Specker \cite{kochen} who demonstrated an inconsistency between the predictions of the QM and the relevant noncontextual hidden variable model.

Lat us first encapsulate the essence of noncontextuality assumption in QM and in a realist hidden variable model. Let an observable $\widehat{A}$ be commuting with $\widehat{B}$ and $\widehat{C}$, while $\widehat{B}$ and $\widehat{C}$ being non-commuting. Given a wave function, QM ensures that the measured statistics of $\widehat{A}$ is independent of, whether the measurement is performed together with  $\widehat{B}$ or $\widehat{C}$. This feature is considered as noncontextuality at the level of QM statistical results. The assumption of `noncontextuality' in any realist model assumes that the predetermined individual outcome of any relevant observable, for a given $\lambda$, is the \emph{same}, whatever be the compatible way that observable is measured. In a non-contextual model, let $v(A)$ be the individual measured values  of $\widehat{A}$, as specified by a $\lambda$,  and let $v(B)$ and $v(C)$ be the individual measured values of the observables $\widehat{B}$ and $\widehat{C}$ respectively, that are also predetermined by the same $\lambda$.  Now, if the individual measured value of the observable is assumed to follow the same context-independence as QM, the crucial question in this regard is, as to what extent the assumption of noncontextuality is compatible with the statistical prediction of QM. It is shown \cite{bell,kochen} that, if the dimension of the Hilbert space is greater than two, the assignment of values by noncontextual realist models is inconsistent with the QM for all possible set of experiments, thereby requiring the hidden variable models to be contextual to reproduce all the predictions of QM. 

Various forms of demonstrations of contextuality have been given by providing a variety of elegant proofs, (see, for example, \cite{ker,cabello18, mermin, peres, cabellosi, ph,  panpm, Kly, yu}) and a flurry of experiments have  been reported \cite{hasegawa1, hasegawa3, nature,liu,ams,zei}. The original proof by Kochen and Specker was demonstrated by using 117 different rays for three-level system, and later simpler versions have been put forward \cite{ker,cabello18,yu} by reducing the number of rays. By using general observables,  Peres \cite{peres} demonstrated an inconsistency between QM and the non-contextual realist models for an entangled state for a pair of two qubits which was later extended to the state-dependent one by  Mermin \cite{mermin}. Using a three-level system, an elegant proof is provided by Klyachko \emph{et al.} \cite{Kly} using only five observables, which has experimentally been tested \cite{zei}. A curious form of contextuality within the formalism of QM has also been demonstrated \cite{ph} without using any realist model. 

Note that, the usual proofs of contextuality are demonstrated by first assuming the non-contextual value assignments by a realist model for a set of compatible observables that are being co-performed and then by showing the contradiction of such value assignments with QM. In other words, such proofs can be regarded as the violation of noncontextual realist model by QM. Recently, for a two-qubit system an interesting non-classical character of QM has been demonstrated by Sala Mayato and Muga \cite{rafasala}. Specifically, the authors have shown that  within the formalism of QM, it is possible to discriminate the different routes that are adopted for measuring a given observable by analysing the reduced density matrices after the measurements. 

In this paper, we provide a simpler proof of such effect than that is given in Ref.\cite{rafasala}. Our argument is based on using a single observable in a qutrit system instead of tensor product observable in four-level system used in Ref.\cite{rafasala}. One may argue that, since different experimental contexts are employed for measuring a given observable that produce two distinct reduced density matrices, such a non-classical effect may be considered as a kind of contextual character inherent in the formalism of QM . Note that, such a form of contextuality is without recourse to any realist hidden variable model. In this demonstration, the degeneracy of the eigenvalues plays an important role and for this, von Neuman projection rule needs to be replaced by L$\ddot{u}$der projection rule. 

The paper is organized as follows. In Section II, we provide a simpler proof of contextuality by discriminating the routes of measuring a given observable for a three-level system. In Section III, we provide a brief sketch about the possible experimental realization of our scheme by explicitly using the probe state. We conclude and discuss our results in Section IV.

\section{'Quantum contextuality' for a single qutrit}
Before presenting our simpler scheme, let us recapitulate the essence of the argument given in Ref.\cite{rafasala}. The use of L$\ddot{u}$der projection rule instead of von Neumann rule is crucial because the eigenvalues of the different observables are degenerate.

Let an observable $\hat{M}$ has discrete eigenvalues $m_1, m_2, m_3,...$ having degeneracies $g_1, g_2, g_3,..$ respectively, consider  $P_{n}^{i}$ is the projection operator associated with each eigenvector and $\rho$ is the initial density matrix of the system. In such a scenario, the von Neumann projection rule prescribes the  reduced density matrix to be
\begin{equation}
\label{von1}
\rho^{\prime}=\sum_{n,i} P_{n}^{i} \rho P_{n}^{i}
\end{equation}
where 
\begin{equation}
P_{n}^{i}=|\chi_{n}^{i}\rangle\langle\chi_{n}^{i}|. 
\end{equation}
However, using the L$\ddot{u}$ders projection rule the reduced density matrix can be obtained as
\begin{equation}
\label{lud1}
\rho^{\prime}=\sum_{n} P_{n} \rho P_{n}
\end{equation}
where 
\begin{align}
\label{lud2}
	P_{n}=\sum_{i=1}^{g_{n}}|\chi_{n}^{i}\rangle\langle\chi_{n}^{i}|
\end{align}
Thus, for an observable with degenerate eigenvalues, the L$\ddot{u}$ders rule provides the reduced density matrix to be partially mixed but the von Neumann rule produces a maximally mixed state. For example, for the measurement of $\hat{I}_{d}$ where degeneracy is $d$, the von Neumann rule provides the reduced density matrix $I_{d}/d$ but according to L$\ddot{u}$der rule the state will remain unchanged, as expected. In contrast to the von Neumann projection rule, the L$\ddot{u}$ders rule does not reduce the state to an eigenstate, it can then be considered a kind of incomplete measurement. The conceptual difference between L$\ddot{u}$der and von Neumann projection rule is discussed in Ref.\cite{heg}.

In Ref.\cite{rafasala}, the authors have considered three observables that were used in Peres-Mermin proof \cite{mermin} of contextuality. Those observables are $\hat{M}_{1}= \hat{\sigma_{x}}\otimes I$, $\hat{M}_{2}= I\otimes\hat{\sigma_{y}}$ and $\hat{M}_{3}=\hat{\sigma_{x}}\otimes\hat{\sigma_{y}}$ where $\hat{M_3}= \hat{M_1}\hat{M_2}$. Since $\hat{M}_{1}$ , $\hat{M}_{2}$ and $\hat{M}_{3}$ are mutually commuting, they have common eigenstates. Now, one can perform the measurement of the observable $\hat{M}_{3}$ by using two different routes; first by directly measuring $\hat{M}_{3}$ and second, by measuring $\hat{M}_{1}$ and then by successively measuring $\hat{M}_{2}$. It has been shown in Ref.\cite{rafasala} that the reduced density matrix obtained  for these two different routes of measurements are \emph{not} the same and thus constitutes a proof of contextuality.      

In this paper, we provide a simpler proof that also uses three commuting observables but in a three-level system instead of four-level system used in  Ref.\cite{rafasala}. In order to demonstrate it, we choose two observables $\hat A$ and $\hat B$ and another observable $\hat{C}$, so that, $\hat{C}= \hat{A}\hat{B}$ is satisfied. The spectral decompositions of the observables $\hat A$, $\hat B$ and $\hat{C}$ are as follows;
\begin{equation}
A= |\phi_1\rangle\langle\phi_1|+|\phi_2\rangle\langle\phi_2|
\end{equation}
whose eigenvalues are $1$, $1$ and $0$ with the corresponding eigenvectors $ |\phi_1\rangle = (1,0,0)^T, |\phi_2\rangle = (0,1,0)^T, |\phi_3\rangle = (0,0,1)^T$ respectively.
\begin{equation}
B= |\phi_2\rangle\langle\phi_2|+|\phi_3\rangle\langle\phi_3|
\end{equation}
whose eigenvalues are $0$, $1$ and $1$ and corresponding eigenvectors are $ |\phi_1\rangle , |\phi_2\rangle$ and $ |\phi_3\rangle $ respectively.
\begin{equation}
C= |\phi_2\rangle\langle\phi_2|
\end{equation}
whose eigenvalues are $0$, $1$ and $0$ and corresponding eigenvectors are $ |\phi_1\rangle , |\phi_2\rangle$ and $ |\phi_3\rangle $ respectively.

Let the initial state of the three-level system is considered to be 
\begin{equation}
\label{inisys}
|\eta\rangle = \alpha|{\phi_1}\rangle + \beta|{\phi_2}\rangle +\gamma|{\phi_3}\rangle
\end{equation}
where $\alpha, \beta, \gamma $ in general complex satisfying $|\alpha|^2+|\beta|^2+|\gamma|^2=1$. 

If the observable $\hat{C}$ is directly measured, the reduce density matrix $\rho_C$ can be obtained by using L$\ddot{u}$der rule given by Eq.(\ref{lud1}), is given by
\begin{eqnarray}
\label{rhoc}
\rho_{C}&=& (|\phi_1\rangle\langle\phi_1|+|\phi_3\rangle\langle\phi_3|)\rho(|\phi_1\rangle\langle\phi_1|+|\phi_3\rangle\langle\phi_3|)+(|\phi_2\rangle\langle\phi_2|)\rho|(\phi_2\rangle\langle\phi_2|) 
\nonumber
\\
&=&|\alpha|^{2}|\phi_1\rangle\langle\phi_1|+|\beta|^{2}|\phi_2\rangle\langle\phi_2| + |\gamma|^{2}|\phi_3\rangle\langle\phi_3|+\alpha^{*}\gamma|\phi_1\rangle\langle\phi_3|+\gamma^{*}\alpha|\phi_3\rangle\langle\phi_1|
\end{eqnarray} 
 
Now, if the observable $\hat{A}$ is measured first and one employs the L$\ddot{u}$der projection rule before measuring the observable $\hat{B}$, the reduce density matrix can be obtained as 
\begin{eqnarray}
\rho_{A}&=& (|\phi_1\rangle\langle\phi_1|+|\phi_2\rangle\langle\phi_2|)\rho(|\phi_1\rangle\langle\phi_1|+|\phi_2\rangle\langle\phi_2|)+(|\phi_3\rangle\langle\phi_3|)\rho|(\phi_3\rangle\langle\phi_3|) \nonumber
\\
&=&  |\alpha|^{2}|\phi_1\rangle\langle\phi_1|+|\beta|^{2}|\phi_2\rangle\langle\phi_2|+|\gamma|^{2}|\phi_3\rangle\langle\phi_3|+\alpha \beta^{\ast} |\phi_1\rangle\langle\phi_2|+\beta\alpha^{\ast}|\phi_2\rangle\langle\phi_1|
\end{eqnarray}
Using $\rho_A$ as initial state, if the observable $\hat B$ is measured the reduce state $\rho_{AB}$ can be obtained by again using L$\ddot{u}$der projection rule is given by
\begin{eqnarray}
\rho_{AB}&=& (|\phi_2\rangle\langle\phi_2|+|\phi_3\rangle\langle\phi_3|)\rho_{A}(|\phi_2\rangle\langle\phi_2|+|\phi_3\rangle\langle\phi_3|)+(|\phi_1\rangle\langle\phi_1|)\rho_{A}|(\phi_1\rangle\langle\phi_1|)
\nonumber
\\
&=&|\alpha|^{2}|\phi_1\rangle\langle\phi_1|+|\beta|^{2}|\phi_2\rangle\langle\phi_2|+|\gamma|^{2}|\phi_3\rangle\langle\phi_3|
\end{eqnarray}
Similarly, if another route is taken by first measuring the observable $\hat{B}$ followed by $\hat{A}$, the reduce density matrix  $\rho_{BA}$ is given by
\begin{equation}
\rho_{BA}= |\alpha|^{2}|\phi_1\rangle\langle\phi_1|+|\beta|^{2}|\phi_2\rangle\langle\phi_2|+|\gamma|^{2}|\phi_3\rangle\langle\phi_3|
\end{equation}

Since $\rho_{BA}\neq \rho_{C}$ and $\rho_{AB}\neq \rho_{C}$, it is in principle possible to distinguish the routes of the measurements of the  observable $\widehat{C}$. 

We thus provided a proof of quantum contextuality for three-level system in terms of discriminating the routes of measuring a given observable. Note that, this demonstration provides a true form of quantum contextuality without reference to any realist hidden variable model. The proof presented here is much simpler than the earlier one \cite{rafasala} which uses entangled state and tensor product observables in a two-qubit system.  

Note here that, by using von Neumann projection rule given by Eq. (\ref{von1}), one obtains, $\rho_{C} = \rho_{AB} = \rho_{BA} = |\alpha|^{2}|\phi_1\rangle\langle\phi_1|+|\beta|^{2}|\phi_2\rangle\langle\phi_2|+|\gamma|^{2}|\phi_3\rangle\langle\phi_3|$ and dose not constitute any proof of contextuality.

We would also like to point out that the degeneracy of the eigenvalues plays a crucial role in our proof. If the observables $\hat{C}$, $\hat{A}$ and $\hat{B}$ all are non-degenerate no such proof of contextuality can be shown. To give an example, we consider three mutually commuting observable  $\widehat{D}_1$, $\widehat{D}_2$,and $\widehat{D}_3$, such that, all have non-degenerate eigenvalues and $\widehat{D}_3 = \widehat{D}_1.\widehat{D}_2$ are as follows;
\begin{equation}
\widehat{D}_1 = (1+\sqrt{3})|d_1\rangle\langle d_1|+(1-\sqrt{3})|d_2\rangle\langle d_2|
\end{equation}
whose eigenvalues are $(1+\sqrt{3})$,$(1-\sqrt{3})$ and $0$ and corresponding unnormalized eigenvectors are $|d_1\rangle = (1, -1+\sqrt{3}, 1)^T$,
$|d_2\rangle = (1, -1-\sqrt{3}, 1)^T$ and $ |d_3\rangle = (-1,0,1)^T$.
 \begin{equation}
\widehat{D}_2 = \sqrt{3}|d_1\rangle\langle d_1|-\sqrt{3}|d_2\rangle\langle d_2|+|d_3 \rangle\langle d_3|
\end{equation}
whose eigenvalues  are $ -\sqrt{3} $,$ \sqrt{3} $ and $1$ and the corresponding unnormalised eigenvectors are $|d_2\rangle$, $|d_1\rangle$ and $ |d_3\rangle$. The eigenvalues of  $\widehat{D}_3$ are $3+\sqrt{3}$, $3-\sqrt3$, $0$ and corresponding eigenvectors are $|d_1\rangle$, $|d_2 \rangle$ and $|d_3 \rangle$.

If one takes an initial state of the form given by
\begin{equation}
|\gamma^{\prime}\rangle= a_1|d_1\rangle + a_2|d_2\rangle + a_3|d_3\rangle
\end{equation}
where $a_1$, $a_2$ and $a_3$ are in general complex. Now following the prescription used before, we can calculate the reduced density matrices $ \rho_{D_{1}D_{2}}$, $ \rho_{D_{2}D_{1}}$ and $ \rho_{D_{3}}$ and show that
\begin{eqnarray}
 \rho_{D_{1}D_{2}} &=& \rho_{D_{2}D_{1}}= \rho_{D_{3}}
\nonumber
\\
&=& |a_1|^2|d_1 \rangle\langle d_1|+ |a_2|^2|d_2 \rangle\langle d_2|+ |a_3|^2|d_3\rangle\langle d_3|
\end{eqnarray}
 which implies that the degeneracy plays the crucial role for showing the contextuality in the earlier case.
 
Note that, the above calculation is done by considering the system state only. In the next section, by introducing the relevant probe states we provide a brief sketch to show that how the non-classical effect we have demonstrated can experimentally realized.
\section{Possible experimental realization of the non-classical effect}
In any measurement scenario the probe state plays the important role, because by looking at the probe pointer position the value of the system is inferred. We provide a measurement model by explicitly using the probe state. In an ideal quantum measurement scenario, after the measurement interaction, the system and probe states become entangled, so that, a perfect one-to-one correspondence between system and probe states is established. But, the measurement may not be completed without mentioning how the observable probabilities can be obtained from such an entangled state, thereby requiring the final step (non-unitary) of quantum measurement - a notion, widely known as collapse of the wave function. For a measured system observable having  degenerate eigenvalues the final step is  to use L$\ddot{u}$der projection rule, otherwise the von Neumann rule is sufficient. Note that inclusion of probe state should retain the same non-classical effect demonstrated above.

Let us again consider the initial system state is given by Eq.(\ref{inisys}) and assume the initial probe state is $|\psi\rangle$, so that, the total state is $|\Psi\rangle=|\psi\rangle|\eta\rangle$. The system-probe entangled state after a suitable measurement interaction for measuring $\hat A$ is given by
\begin{equation}
|\Psi_{A}\rangle = \alpha|\psi_1\rangle|\phi_1\rangle+\beta|\psi_1\rangle|\phi_2\rangle+\gamma|\psi_0\rangle|\phi_3\rangle
\label{psia}
\end{equation}
where $|\psi_{0}\rangle$ and $|\psi_{1}\rangle$ are the post-interaction probe states corresponding to the eigenvalues $1$ and $0$ respectively. For ideal measurement situation $\langle\psi_{1}|\psi_{0}\rangle=0$ needs to be satisfied. The entangled state given by Eq.(\ref{psia}) is then subject to measurement of the observable $\hat{B}$. After this measurement interaction, a different system-apparatus entangled state is obtained is given by 
\begin{equation}
|\Psi_{AB}\rangle = \alpha|\psi_{10}\rangle|\phi_1\rangle+\beta|\psi_{11}\rangle|\phi_2\rangle+\gamma|\psi_{01}\rangle|\phi_3\rangle
\label{psiab}
\end{equation}
where $|\psi_{10}\rangle$ is the post-interaction probe state corresponding to the $0$ eigenvalue of the observable $\hat{B}$. Again, for an ideal measurement, $\langle\psi_{10}|\psi_{11}\rangle=0$ and $\langle\psi_{10}|\psi_{01}\rangle =0$.

Next, the order of measurement is swapped by first considering the interaction for measuring $\hat{B}$ followed by $\hat{A}$. The system-probe entangled state after the measurement of $\hat{B}$ can be written as
\begin{equation}
|\Psi_{B}\rangle = \alpha|\psi_0\rangle|\phi_1\rangle+\beta|\psi_1\rangle|\phi_2\rangle+\gamma|\psi_1\rangle|\phi_3\rangle
\label{psib}
\end{equation}
Now measuring $\hat{A}$ using $|\Psi_{B}\rangle$ as initial state, one obtains
\begin{equation}
|\Psi_{BA}\rangle = \alpha|\psi_{01}\rangle|\phi_1\rangle+\beta|\psi_{11}\rangle|\phi_2\rangle+\gamma|\psi_{10}\rangle|\phi_3\rangle
\label{psiba}
\end{equation}
If the observable $\hat{C}$ is directly measured, the system-apparatus entangled state is given by  
\begin{equation}
|\Psi_{C}\rangle = \alpha|\psi_0\rangle|\phi_1\rangle+\beta|\psi_1\rangle|\phi_2\rangle+\gamma|\psi_0\rangle|\phi_3\rangle
\label{psic}
\end{equation}

Now, the observable probe signals can be obtained by suitably using the L$\ddot{u}$ders projection rule in the Eqs.(\ref{psiab}), (\ref{psiba}) and (\ref{psic}) corresponding to the three routes of the measurements. 

One may also check that inclusion of probe state does not effect the contextuality argument presented in the above section. For this, by taking the partial trace over the probe state, the reduced density matrix $\rho^{\prime}_{AB}= Tr_{p}[|\Psi_{AB}\rangle\langle\Psi_{AB}|]$ of the system for the first route is equal to $\rho_{AB}$. Similarly, it can be shown that $\rho_{C}^{\prime}= Tr_{p}[|\Psi_{C}\rangle\langle\Psi_{C}|]=\rho_{C}$. 
\section{summary and conclusions}
John Bell \cite{bell} had remarked that ``The result of an observation may reasonably depend not only on the state of the system .... but also on the complete disposition of the apparatus". Along the same vain, we provided a curious form of nonclassical effect by showing that the results of the measurement of a particular observable may reasonably depend not only on the state but also on the contexts in which the observable is being measured.  Specifically, we have shown that the reduced density matrices that are produced by employing the different routes for measuring a given observable can be empirically distinguished. Such a nonclassical effect can then be considered as a contextuality within the formalism of QM. This form of contextuality was first demonstrated in Ref.\cite{rafasala} for  a four-level system. In this paper, we demonstrated a simpler proof for a qutrit system which is also within the formalism of QM without recourse to any realist hidden variable model. Note that, any proof of quantum violation of noncontextual realist model requires at least four observables. Here, we use a single observable but employed two different routes to measure it. As regards the relation to our proof of contextuality with the existing usual proof in terms of the violation of non-contextual realist model, it is trivial that any realist hidden variable model which reproduces the effect of non-classicality demonstrated here should be contextual.   
\section*{Acknowledgments}
AKP thanks Prof. R. Sala Mayato for helpful discussions. AKP acknowledges the support from Ramanujan Fellowship research grant. KM gratefully acknowledges the summer student fellowship from TEQIP grant and the local hospitality from NIT Patna. 

\end{document}